%% file: main.tex
\documentclass[conference]{IEEEtran}

\IEEEoverridecommandlockouts

\usepackage{amsmath,amsthm,relsize}
\usepackage{amsfonts, amssymb, cuted}
\usepackage{cleveref}
\usepackage{mathtools}
\usepackage{algorithm}
\usepackage[noend]{algpseudocode}
\usepackage{cite}
\usepackage{xcolor}
\usepackage{soul}
\usepackage{graphicx}
\usepackage{tikz}
\usepackage{pgfplotstable}
\usepackage{pgfplots}
\usepackage{nicefrac}
\usepackage{enumitem}
\usepackage{support-caption}
\usepackage{subcaption}
\usepackage[font=footnotesize]{caption}
\usepackage{multirow}
\pgfplotsset{compat=1.17}
\usepackage{acro}
\usepackage{pifont}

\input{MyCommands}

\newcommand{\subparagraph}{}
\usepackage{titlesec}
\titlespacing\section{3pt}{6pt plus 4pt minus 2pt}{6pt plus 2pt minus 2pt}
\titlespacing\subsection{3pt}{4pt plus 4pt minus 2pt}{4pt plus 2pt minus 2pt}
\titlespacing\subsubsection{3pt}{3pt plus 4pt minus 2pt}{0pt plus 2pt minus 3pt}

\title{Multi-antenna Coded Caching at Finite-SNR:\\ Breaking Down the Gain Structure}

\begin{document}

\author{\IEEEauthorblockN{MohammadJavad Salehi and Antti T\"olli} \\
\IEEEauthorblockA{
    Centre for Wireless Communications, University of Oulu, 90570 Oulu, Finland \\
    \textrm{E-mail: \{firstname.lastname\}@oulu.fi}
    }
\thanks{This work is supported by the Academy of Finland under grants no. 346208 (6G Flagship), 319059 (CCCWEE), and 343586 (CAMAIDE), and by the Finnish Research Impact Foundation under the project 3D-WIDE.}
}

\maketitle

\begin{abstract}
Multi-antenna coded caching (CC) techniques are considered viable options for achieving higher data rates in future networks, especially for the prominent use case of multimedia-driven applications. However, despite their information-theoretic analyses, which are thoroughly studied in the literature, the research on the finite-SNR performance of multi-antenna CC techniques is not yet mature. In this paper, we try bridging this gap by breaking down, categorizing, and studying the effect of six crucial parameters affecting the finite-SNR performance of multi-antenna CC schemes. We also investigate the interaction of different parameters and clarify how they could affect the implementation complexity in terms of the necessary computation and subpacketization. Theoretical discussions are followed and verified by numerical analysis.
\end{abstract}

\begin{IEEEkeywords}
coded caching; multi-antenna communications; multiplexing; multicasting; beamforming
\end{IEEEkeywords}

\section{Introduction}
\label{section:intro}
\input{Intro}

\section{System Model}
\label{section:sysmodel}
\input{SysModel}

\section{MISO-stemmed parameters}
\label{section:misogains}
\input{MISOGains}

\section{CC-stemmed parameters}
\label{section:ccgains}
\input{CCGains}


\section{Conclusion and Future Work}
In this paper, we studied six parameters affecting the finite-SNR performance of coded caching schemes in MISO setups. Out of the six parameters, three stemmed from the multi-antenna transmission part and the rest relied on cache-aided communications. For each parameter, we provided a brief explanation, followed by simulation results and insights.

As a quick summary, both multi-antenna and cache-aided communication techniques play important roles in finite-SNR. However, being DoF-optimal is not always preferred; one can slightly reduce the spatial multiplexing gain and still get better results due to improved beamformer directivity. Deviation from the optimal point also narrows the performance gap of simple zero-force beamformers with more complex optimized ones. On the other hand, CC gain is independent of the SNR regime, bit-level schemes have better performance at finite-SNR due to improved multicasting gain, and designing bit-level schemes without SIC requirement is of interest.

Future research directions include expanding the results to multi-input multi-output (MIMO) communication setups where the users are also equipped with antenna arrays, a more thorough mathematical analysis of the effect of the parameters, and expanding the results considering a more diverse set of baseline CC schemes.

\bibliographystyle{IEEEtran}
\bibliography{references,whitepaper}

\end{document}

%% file: MyCommands.tex
\theoremstyle{definition}

\newcommand{\CA}[0]{{\mathcal{A}}}
\newcommand{\CB}[0]{{\mathcal{B}}}

\newcommand{\CF}[0]{{\mathcal{F}}}

\newcommand{\CR}[0]{{\mathcal{R}}}
\newcommand{\CS}[0]{{\mathcal{S}}}
\newcommand{\CT}[0]{{\mathcal{T}}}
\newcommand{\CU}[0]{{\mathcal{U}}}
\newcommand{\CV}[0]{{\mathcal{V}}}
\newcommand{\CW}[0]{{\mathcal{W}}}

\newcommand{\Bh}[0]{{\mathbf{h}}}

\newcommand{\Bw}[0]{{\mathbf{w}}}
\newcommand{\Bx}[0]{{\mathbf{x}}}

\newcommand{\BH}[0]{{\mathbf{H}}}



%% file: Intro.tex
Wireless communication networks are under mounting pressure to support higher data rates, and this trend will continue by emerging new services such as extended reality (XR) applications~\cite{rajatheva2020whiteb}. 
The coded caching (CC) technique~\cite{maddah2014fundamental} is considered a viable candidate to address this issue. The critical property of CC is that it enables the data storage memory in network devices to be efficiently used as a communication resource, especially for the prominent use case of multimedia-driven applications~\cite{mahmoodi2021non,salehi2022enhancing}.
Interestingly, the caching gain of CC scales with the cumulative cache size of all users in the network~\cite{maddah2014fundamental}, and can be combined with the spatial multiplexing gain of incorporating multiple antennas at the transmitter~\cite{shariatpanahi2016multi,shariatpanahi2018physical} or receivers~\cite{salehi2021MIMO}.
However, this promising CC gain is followed by crucial practical bottlenecks hindering its applicability.
Two important such bottlenecks are the subpacketization issue and the optimized beamformer design (for improved performance at the finite-SNR regime).
Former stems from the fact that the original multi-antenna CC scheme in~\cite{shariatpanahi2018physical} required splitting files into exponentially-growing numbers of smaller parts, and the latter is due to the fact that designing optimized beamformers for this scheme requires solving complex non-convex optimization problems~\cite{tolli2017multi}.

While the subpacketization issue seems to be fundamental in single-antenna systems~\cite{yan2018placement}, multi-antenna setups provide flexibility for reducing the required subpacketization without altering the theoretical degrees-of-freedom~(DoF) gains~\cite{lampiris2018adding,salehi2021low,mohajer2020miso}. The key to achieving this reduction is in using a new CC approach where the cache-aided interference cancellation is done \emph{before} decoding the data and in the signal domain~\cite{lampiris2018adding}. Following~\cite{salehi2022enhancing}, we use the term \emph{signal-level} while referring to CC schemes with this new approach and denote the traditional CC schemes (with cache-aided interference cancellation after decoding the received signal) as \emph{bit-level}. Interestingly, even though signal-level CC schemes were initially developed for subpacketization reduction, the work in~\cite{salehi2021low} showed that they also allow simplifying the optimized beamformer design by altering the underlying multicast structure of bit-level schemes.

However, achieving the same DoF and applicability of optimized beamformers does \emph{not} reveal the whole story about the finite-SNR performance of signal-level CC schemes. For example, it was shown in~\cite{salehi2019subpacketization} that with the same DoF, moving eventually from a pure signal-level approach (smallest subpacketization) to a pure bit-level one (largest subpacketization), the performance is improved at the finite-SNR regime. Also, the work in~\cite{tolli2017multi} revealed the fact that being DoF-optimal is not necessarily preferred in finite-SNR as the DoF can be traded-off with the more effective beamformer directivity parameter. Such results clarify a need to better understand all the parameters affecting the finite-SNR performance of CC schemes and to investigate the possibility of designing new CC schemes and beamforming techniques to better tune the performance for any given complexity constraint.

\begin{figure}[t]
    \centering
    \includegraphics[width=0.79\columnwidth]{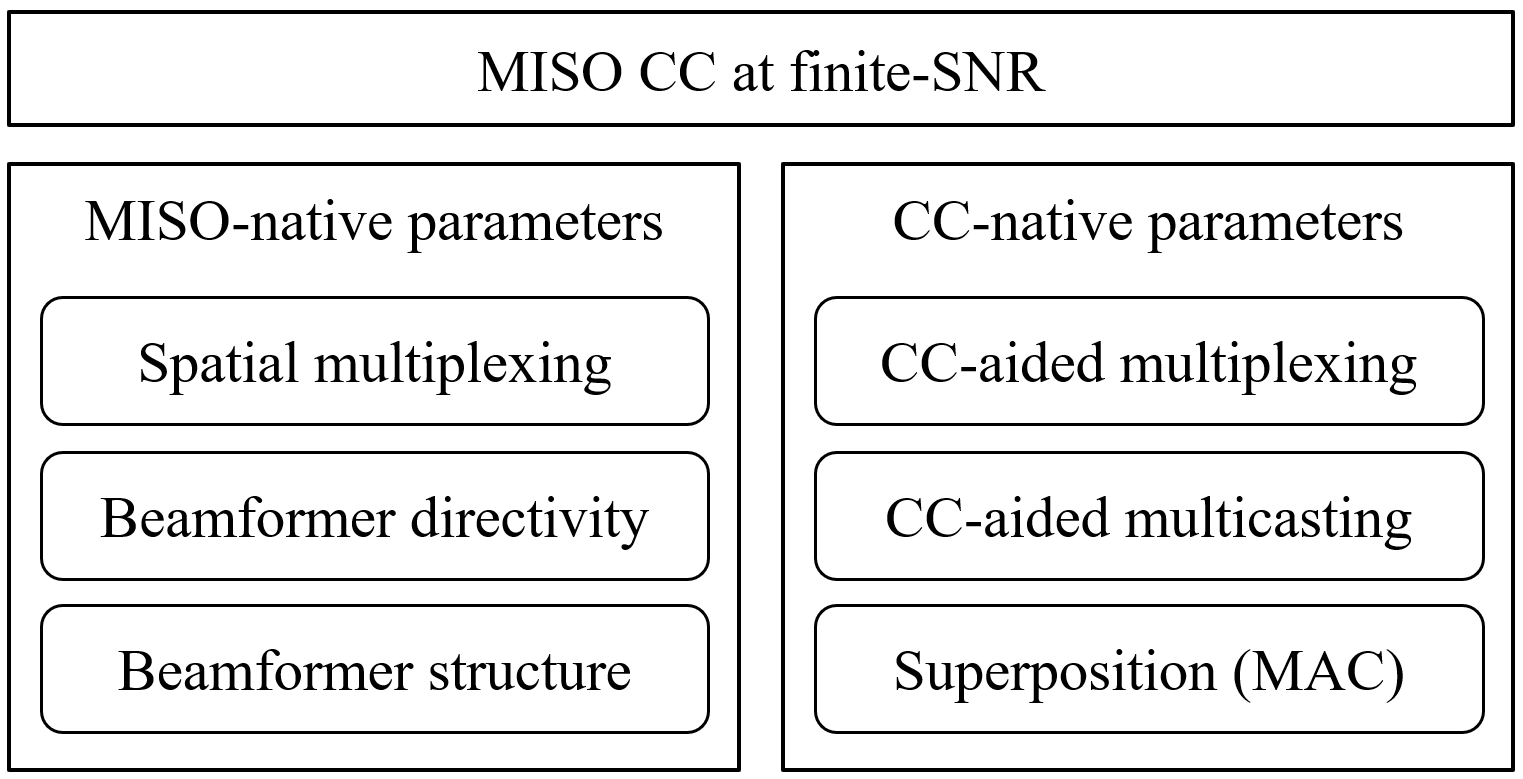}
    \caption{Parameters affecting the finite-SNR performance of MISO CC schemes}
    \label{fig:gain_breakdown}
\end{figure}

This paper breaks down and classifies all the parameters affecting the finite-SNR performance of multi-antenna CC schemes. Specifically, we identify six different parameters, as shown in Figure~\ref{fig:gain_breakdown}, and analyze how they affect the achievable rate of CC schemes in different SNR values. For each parameter, we provide the definition and a brief discussion on how it can be adjusted, followed by theoretical insights and numerical simulation results. As discussed, this paper is not the first to identify all these parameters and study their effects. However, it is the first to gather, organize, and study the interaction of all such performance-affecting parameters. We also carefully discuss how each parameter could affect the implementation complexity and provide guidelines for selecting the appropriate scheme following the available transmission power and acceptable complexity. 

In this paper, we have used the following notations. Bold-face lower- and upper-case letters denote vectors and matrices, respectively. Calligraphic letters are used to represent sets. For an integer $K$, $[K]$ is the set of numbers $\{1,2,\cdots,K\}$. The symbol $\oplus$ denotes the bit-wise XOR operation. For two sets $\CA$ and $\CB$, $\CA \backslash \CB$ is the set of elements of $\CA$ that are not in $\CB$. 
$|\CA|$ is the number of elements in $\CA$. Set brackets and element separators are sometimes dropped for notational simplicity.

%% file: SysModel.tex
We consider a multi-input single-output (MISO) cache-aided communication setup. A single, $L$-antenna transmitter with spatial multiplexing gain of $\alpha \le L$ communicates with $K$ single-antenna users over a shared wireless link. Every user has a cache memory of size $MF$ bits and requests files from a library $\CF$ of $N$ files, each with the size of $F$ bits. The CC gain is defined as $t \equiv \frac{KM}{N}$. For notational simplicity, we use a normalized data unit and drop $F$ in our subsequent notations. We also use \textit{CC}-$(K,t,L,\alpha)$ to refer to this setup.

The system operation consists of two phases, content placement and delivery. In the content placement phase, which is done at the low network traffic time, cache memories of the users are filled up by data chunks of files in $\CF$. 
Then, in the delivery phase, after each user $k \in [K]$ reveals its requested file $W(k) \in \CF$, the server creates and transmits a set of codewords such that all the users can decode their requested files using their cache contents and the received signals.
The goal is to design content placement and delivery phases such that the symmetric rate, defined as the amount of data delivered to all the users over a defined time frame, is maximized. In this paper, we consider only CC schemes with uncoded data placement and single-shot data delivery, for which the MISO-CC scheme in~\cite{shariatpanahi2018physical} is shown to be DoF-optimal~\cite{lampiris2018resolving}.


Data delivery is done using a set of transmission vectors, which are sent, for example, in consecutive time slots. A transmission vector $\Bx_{\CS}$ delivers (parts of) the requested data to every user in the subset $\CS \subseteq [K]$ of users, where $|\CS| = t+\alpha$. This is done through the parallel transmission of codewords $X_{\CT}$, where $\CT \subseteq \CS$. Every codeword $X_{\CT}$ includes useful data for every user $k \in \CT$ and is precoded with a beamforming vector $\Bw_{\CR(\CT)}$ that nulls-out or suppresses the interference of $X_{\CT}$ at every user $k \in \CR(\CT)$. Using $\Bar{\CT}(\CS)$ to denote the set of subsets $\CT \subseteq \CS$ for which a codeword $X_{\CT}$ is built, we can write
%
\begin{equation}
\label{eq:general_transmission_vector}
    \Bx_{\CS} = \sum_{\CT \in \Bar{\CT}(\CS)} X_{\CT} \Bw_{\CR(\CT)} .
\end{equation}
Then, after the transmission of $\Bx_{\CS}$, the received signal at user $k \in \CS$ can be modeled as
\begin{equation}
\label{eq:general_received_signal}
    y_{\CS} (k) = \sum_{\CT \in \Bar{\CT}(\CS)} X_{\CT} {\Bh_{\CS}(k)}^H \Bw_{\CR(\CT)} + z_{\CS} (k),
\end{equation}
where $\Bh_{\CS}(k) \in \mathbb{C}^{L \times 1}$ and $z_{\CS} (k)$ represent the channel vector and noise at user $k$ during the transmission of $\Bx_{\CS}$. In this paper, we use assume $\Bh_{\CS}(k)$ does not change during the transmission of $\Bx_{\CS}$, and full channel state information (CSI) is available at the transmitter. We also use $\Bh_k \equiv \Bh_{\CS}(k)$ and $z_k \equiv z_{\CS}(k)$ for notational simplicity.

Let us denote the maximum time required for every user $k$ in $\CS$ to decode $y_{\CS}(k)$ as $T_{\CS}$. Then, the total delivery time is $T = \sum_{\CS} T_{\CS}$, and the symmetric channel rate, used as the comparison metric here, can be defined as
\begin{equation}
\label{eq:sym_rate}
    R_{\mathrm{sym}} = \frac{K-t}{T}.
\end{equation}
Note that this metric reflects how well the wireless communication channel is used and excludes the effect of the local caching gain. This is unlike the total symmetric rate $\frac{K}{T}$ widely used in the literature~\cite{shariatpanahi2018physical}. Of course, the two metrics can be converted into each other with a simple coefficient.

%% file: MISOGains.tex
As shown in Figure~\ref{fig:gain_breakdown}, we study three MISO-stemmed parameters affecting the finite-SNR performance of coded caching schemes. For this study, we choose the MISO scheme in~\cite{shariatpanahi2018physical} as the baseline. This is because this scheme enables the maximum performance for all CC-stemmed parameters and hence, provides a fair structure for capturing the effect of the MISO-stemmed parameters considered here.

With the MISO scheme in~\cite{shariatpanahi2018physical} as the baseline, during the placement phase, we split every file $W \in \CF$ into $\binom{K}{t}$ subfiles $W_{\CU}$, where $\CU \subseteq [K]$ and $|\CU| = t$. Then, at the cache memory of user $k \in [K]$, we store $W_{\CU}$ for every $W \in \CF$ and $\CU \ni k$. Subsequently, during the delivery phase, we create a separate transmission vector $\Bx_{\CS}$ for \emph{every} subset of users $\CS \subseteq [K]$ with size $|\CS| = t+\alpha$. Given one such set $\CS$, the transmission vector $\Bx_{\CS}$ includes a codeword for \emph{every} user subset of $\CS$ with $t+1$ users, i.e., $\Bar{\CT}(\CS) = \{\CT \subseteq \CS \; \mid \; |\CT| = t+1\}$.\footnote{It should be noted that during the delivery phase, we should further split every subfile $W_{\CU}$ into $Q = \binom{K-t-1}{\alpha-1}$ smaller subpackets $W_{\CU}^q$, $q \in [Q]$, to ensure new data chunks are delivered during every transmission. In this paper, we have ignored noting this second-level subpacketization as it does not affect the discussions. The effects are considered in numerical simulations, though.} 

For better clarification, let us consider an example network of \textit{CC}-$(6,2,4,\alpha)$.
During the placement phase, we split each file in $\CF$ into $\binom{6}{2} = 15$ subfiles and store five of them in the cache memory of each user. For example, a file $A \in \CF$ is split into subfiles $A_{12}$, $A_{13}$, $\cdots$, $A_{56}$,
from which $A_{12}$, $A_{13}$, $A_{14}$, $A_{15}$, and $A_{16}$ (i.e., all subfiles $A_{\CU}$ for which $1 \in \CU$) are stored in the cache memory of user 1.
Now, we review the delivery phase, assuming $\alpha = 1,4$. For notational simplicity, let us denote the files requested by users 1-6 as $A,B,\cdots,F$, respectively.

\noindent $\bullet$ $\alpha = 1$. In this case, we have to create $\binom{K}{t+\alpha} = \binom{6}{3} = 20$ transmission vectors $\Bx_{\CS}$. We also have $\Bar{\CT}(\CS) = \{\CS\}$, i.e., every vector $\Bx_{\CS}$ includes only a single codeword $X_{\CT}$. As a result, there is no need to null-out (or suppress) the interference from $X_{\CT}$ at other users, and $\CR(\CT) = \varnothing$. For example, considering $\CS = \{1,2,3\}$, the transmission vector $\Bx_{123}$ is built as
\begin{equation}
    \begin{aligned}
    \Bx_{123} = X_{123} \Bw_{\varnothing} ,
    \end{aligned}
\end{equation}
where the codeword $X_{123} = A_{23} \oplus B_{13} \oplus C_{12}$ includes data for all users 1, 2, and 3. 

Now, let us consider the decoding process at user~1 after transmitting $\Bx_{123}$. Using~\eqref{eq:general_received_signal}, this user receives
\begin{equation}
    y_{123}(1) = X_{123} \Bh_1^T \Bw_{\varnothing} + z_1 ,
\end{equation}
and can extract $X_{123}$ with a maximum rate of
\begin{equation}
    r_{123}(1) \le \log \big(1 + \frac{|\Bh_1^T \Bw_{\varnothing}|}{N_0} \big) .
\end{equation}
Then, user~1 has to use its cached contents to remove unwanted data terms $B_{23}$ and $C_{12}$ from $X_{123}$. Similarly, users~2 and~3 can also get their requested data terms.

\noindent $\bullet$ $\alpha = 4$. In this case, we create only a single transmission vector $\Bx_{1\cdots 6}$. However, this transmission vector is comprised of $\binom{6}{3} = 20$ codewords, where each codeword delivers data to three users and its interference is suppressed by beamforming at the other three users. In other words,
\begin{equation}
\label{eq:bit_level_exmp}
\begin{aligned}
    \Bx_{1\cdots 6} = X_{123} \Bw_{456} + X_{124} \Bw_{356} + \cdots + X_{456} \Bw_{123}.
\end{aligned}
\end{equation}
The codewords are similar to the $\alpha = 1$ case. For example, we have $X_{123} = A_{23} \oplus B_{13} \oplus C_{12}$, $X_{124} = A_{24} \oplus B_{14} \oplus D_{12}$, and $X_{456} = D_{56} \oplus E_{46} \oplus F_{45}$.

Now, let us consider the decoding process at user~1. The received signal at this user is
\begin{equation}
\begin{aligned}
    y_{1\cdots 6}(1) = X_{123} \Bh_1^T \Bw_{456} + \cdots + X_{456} \Bh_1^T \Bw_{123} + z_1 .
\end{aligned}
\end{equation}
Unlike the $\alpha=1$ case, user~1 now has to decode its requested data from a multiple access channel (MAC) of size~10, using a more complex successive interference cancellation (SIC) receiver, as there exist $\binom{5}{2} = 10$ codewords $X_{\CT}$ for which $1 \in \CT$. The maximum decoding rate in this case is determined by a rate region. Assume $\hat{\CV}(1)$ and $\Bar{\CV}(1)$ include the user subsets $\CT \subseteq \CS$ of size $t+1=3$, that include and not include user~1, respectively. In other words
\begin{align}
    \hat{\CV}(1) &= \{123, 124, \cdots, 156 \} , \\
    \Bar{\CV}(1) &= \{234, 235, \cdots, 456 \} .
\end{align}
Then, the SINR term $\gamma_{\hat{\CV}}$ for a desired term $A_{\hat{\CV}}$, $\hat{\CV} \in \hat{\CV}(1)$, is calculated as
\begin{equation}
    \gamma_{\hat{\CV}} = \frac{|\Bh_1^T\Bw_{[6] \backslash \hat{\CV} }|^2}{\sum_{\Bar{\CV} \in \Bar{\CV}(1)} |\Bh_1^T \Bw_{[6] \backslash \Bar{\CV}}|^2 + N_0},
\end{equation}
and the decoding rate $r_{1\cdots 6}(1)$ satisfies
\begin{equation}
    r_{1\cdots 6}(1) \le \! \frac{1}{|\CW|}  \log \! \big( 1 \! +  \!\!\! \sum_{\hat{\CV} \in \CW} \gamma_{\hat{\CV}} \big), \;\forall \CW \subseteq \hat{\CV}(1), |\CW| > 0.
\end{equation}
Finding optimal beamformers maximizing the rate in this case is difficult as it requires solving a non-convex optimization problem with the number of constraints growing exponentially with the MAC size~\cite{tolli2017multi}. We will discuss the effect of different beamforming strategies in Section~\ref{section:beamformer_type}.

\subsection{Spatial multiplexing effect}
Spatial multiplexing gain $\alpha \le L$ determines the number of users at which we can null-out or suppress the interference caused by each term. 
%
As a larger $\alpha$ means serving more users in parallel (and hence, a larger DoF), most works in the literature have assumed $\alpha = L$. 
However, as discussed in~\cite{tolli2017multi}, the DoF is not the best metric at the finite-SNR regime; setting $\alpha < L$ results in better performance due to an increased \emph{beamformer directivity} gain.

In this section, to have a fair analysis of the effect of the spatial multiplexing gain, we set $\alpha = L$ to remove the effect of beamformer directivity. We perform numerical simulations for \textit{CC}-$(6,2,L,L)$ setup, 
where $L \in \{1,2,3,4 \}$. Simulation results are shown in Figure~\ref{fig:smux_plot} and Table~\ref{tab:smux_bar}. In all simulations, we use optimized beamformers for maximizing performance. Highlights of the results are:
\begin{enumerate}
    \item If $\alpha = L$, a larger spatial multiplexing gain always improves the performance. This is because the DoF is increased, and beamformer directivity has no effect;
    \item Performance improvement (over the $\alpha=L=1$ case) is more prominent in the finite-SNR regime. This is because the $\alpha = L = 1$ case represents omni-casting data in all directions, which is very inefficient in finite-SNR;
    \item Using multi-antenna transmission techniques is important in finite-SNR, even when they are used only for spatial multiplexing and not to improve beamformer directivity.
\end{enumerate}

\input{Figs/SMUX-Plot}

\begin{table}[t]
    \centering
    \begin{tabular}{c||c|c|c|c|c|c|c}
         SNR ($\mathrm{dB}$) & 0 & 5 & 10 & 15 & 20 & 25 & 30  \\
         \hline \hline
         $L=4$ & 405 & 297 & 212 & 164 & 140 & 130 & 124 \\
         \hline
         $L=3$ & 317 & 231 & 164 & 126 & 107 & 98 & 92 \\
         \hline
         $L=2$ & 138 & 109 & 77 & 55 & 45 & 41 & 40
    \end{tabular}
    \caption{Rate benefit (\%) over $L=1$ case, $K=6$, $t=2$, $\alpha=L$}
    \label{tab:smux_bar}
\end{table}

\subsection{Beamformer directivity effect}
As discussed, although choosing $\alpha < L$ decreases the DoF, it may improve the finite-SNR performance by enhancing beamformer directivity. In fact, as we decrease $\alpha$ (compared with $L$), the ratio of the variables to constraints in the beamformer optimization problem grows larger, enabling designing narrower beams that better direct data signals to end users~\cite{tolli2017multi}, thus improving the performance, especially in finite-SNR.

To investigate the effect of beamforming directivity, we have provided simulation results for a \textit{CC}-$(6,2,4,\alpha)$ setup, where $\alpha \in \{1,2,3,4\}$, in Figure~\ref{fig:sbf_plot} and Table~\ref{tab:sbf_bar}.
Again, optimized beamformers are used to maximize the performance. Highlights of the results are:
\begin{enumerate}
    \item Beamformer directivity has a very strong effect in finite-SNR. It can even fully compensate for the performance loss due to decreased DoF. These results comply with~\cite{tolli2017multi};
    \item The best choice seems to be choosing $\alpha$ to be slightly smaller than $L$. This choice also slightly simplifies the beamformer design problem. Of course, in smaller SNR values (the negative range, which is not considered here), this recommendation may change.
\end{enumerate}

\input{Figs/SBF-Plot}

\begin{table}[t]
    \centering
    \begin{tabular}{c||c|c|c|c|c|c|c}
         SNR ($\mathrm{dB}$) & 0 & 5 & 10 & 15 & 20 & 25 & 30  \\
         \hline \hline
         $\alpha=4$ & -1 & 7 & 15 & 25 & 36 & 47 & 56 \\
         \hline
         $\alpha=3$ & 7 & 13 & 20 & 29 & 37 & 43 & 47 \\
         \hline
         $\alpha=2$ & 5 & 10 & 15 & 20 & 23 & 25 & 26
    \end{tabular}
    \caption{Rate benefit (\%) over $\alpha=1$ case, $K=6$, $t=2$, $L=4$}
    \label{tab:sbf_bar}
\end{table}

\subsection{Beamformer structure effect}
\label{section:beamformer_type}
The beamforming vectors $\Bw_{\CR(\CT)}$ in~\eqref{eq:general_transmission_vector} can be designed in different ways. The simplest strategy is zero-forcing, i.e., to design $\Bw_{\CR(\CT)}$ such that $\Bh_k^T \Bw_{\CR(\CT)} = 0$ for every user $k \in \CR(\CT)$. In other words, $\Bw_{\CR(\CT)}$ should lie in the null-space of the matrix $\BH = [\Bh_{k_1},...,\Bh_{k_{\alpha-1}}]$ formed by concatenating channel vectors of users in set $\CR(\CT) = \{k_1,...,k_{\alpha-1}\}$. This is straightforward if $\alpha = L$, as in this case, the null-space is of dimension one. However, if $\alpha < L$, the null-space has higher dimensions and the beamformer vector can be any vector in that space. In this paper, we assume the best vector (for maximizing the symmetric rate) is found in the null-space by solving an optimization problem. The details are removed due to the lack of space.

Of course, zero-forcing is not the optimum strategy in the finite-SNR regime~\cite{tolli2017multi}. Instead, one needs to use optimized beamformers by maximizing the symmetric rate given the SNR constraints. Indeed, this can result in non-convex optimization problems, necessitating solutions such as successive convex approximation (SCA)~\cite{tolli2017multi}. Of course, the underlying scheme can be tweaked to reduce the optimized beamformer design complexity (e.g., using the signal-level scheme in~\cite{salehi2020lowcomplexity}). However, we still use the baseline MISO scheme in~\cite{shariatpanahi2018physical} here, as it maximizes the positive effects of all CC-stemmed parameters.

To investigate the effect of the beamformer structure, we have provided numerical simulation results for zero-force and optimized beamforming strategies for a \textit{CC}-$(6,2,4,\alpha)$ setup, $\alpha \in \{2,4\}$, in Figure~\ref{fig:stype_plot} and Table~\ref{tab:stype_bar}. 
Highlights of the results are:
\begin{enumerate}
    \item Performance gap of the two strategies is very big in low-SNR but narrows down as the SNR grows;
    \item Performance gap at low-SNR gets smaller if $\alpha < L$. This is because with $\alpha < L$, zero-force beamformers are selected from a larger null-space~\cite{tolli2017multi};
    \item If computation capability is a bottleneck, choosing $\alpha < L$ and zero-force beamforming is a wise choice for finite-SNR communications, as we also benefit from improved beamformer directivity.
\end{enumerate}

\input{Figs/STYPE-Plot}

\begin{table}[t]
    \centering
    \begin{tabular}{c||c|c|c|c|c|c|c}
         SNR ($\mathrm{dB}$) & 0 & 5 & 10 & 15 & 20 & 25 & 30  \\
         \hline \hline
         $\alpha=4$ & 582 & 347 & 167 & 69 & 27 & 9 & 2 \\
         \hline
         $\alpha=2$ & 81 & 47 & 26 & 16 & 11 & 8 & 7 \\
    \end{tabular}
    \caption{Rate benefit (\%) over ZF beamforming, $K=6$, $t=2$, $L=4$}
    \label{tab:stype_bar}
\end{table}

%% file: Figs/SMUX-Plot.tex
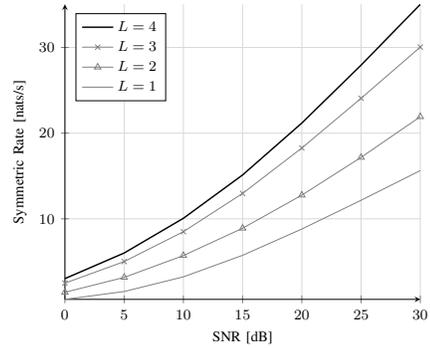
\begin{figure}[t]
    \centering
    \resizebox{0.65\columnwidth}{!}{%
    
    \begin{tikzpicture}

    \begin{axis}
    [
    axis lines = left,
    xlabel = \smaller {SNR [dB]},
    ylabel = \smaller {Symmetric Rate [nats/s]},
    ylabel near ticks,
    xmax=30,
    legend pos = north west,
    ticklabel style={font=\smaller},
    grid=both,
    major grid style={line width=.2pt,draw=gray!30},
    ]
    
    \addplot[black,thick]
    table[y=L4OPT,x=SNR]{Data/SMUX.tex};
    \addlegendentry{\smaller $L=4$}
    
    \addplot[mark = x,gray]
    table[y=L3OPT,x=SNR]{Data/SMUX.tex};
    \addlegendentry{\smaller $L=3$}
    
    \addplot[mark = triangle,gray]
    table[y=L2OPT,x=SNR]{Data/SMUX.tex};
    \addlegendentry{\smaller $L=2$}
    
    \addplot[gray]
    table[y=L1OPT,x=SNR]{Data/SMUX.tex};
    \addlegendentry{\smaller $L=1$}

    \end{axis}

    \end{tikzpicture}
    }
    \caption{Spatial multiplexing effect, $K=6$, $t=2$, $\alpha=L$}
    \label{fig:smux_plot}
\end{figure}

%% file: Figs/SBF-Plot.tex
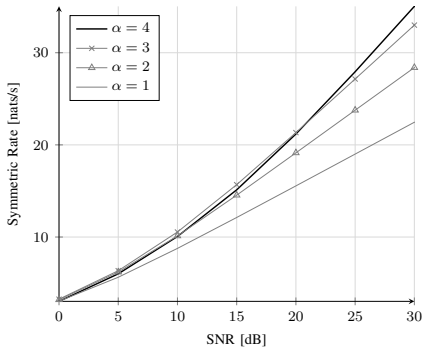
\begin{figure}[t]
    \centering
    \resizebox{0.65\columnwidth}{!}{%
    
    \begin{tikzpicture}

    \begin{axis}
    [
    axis lines = left,
    xlabel = \smaller {SNR [dB]},
    ylabel = \smaller {Symmetric Rate [nats/s]},
    ylabel near ticks,
    xmax=30,
    legend pos = north west,
    ticklabel style={font=\smaller},
    grid=both,
    major grid style={line width=.2pt,draw=gray!30},
    ]
    
    \addplot[black,thick]
    table[y=A4OPT,x=SNR]{Data/SBF.tex};
    \addlegendentry{\smaller $\alpha = 4$}
    
    \addplot[mark = x,gray]
    table[y=A3OPT,x=SNR]{Data/SBF.tex};
    \addlegendentry{\smaller $\alpha = 3$}
    
    \addplot[mark = triangle,gray]
    table[y=A2OPT,x=SNR]{Data/SBF.tex};
    \addlegendentry{\smaller $\alpha = 2$}
    
    \addplot[gray]
    table[y=A1OPT,x=SNR]{Data/SBF.tex};
    \addlegendentry{\smaller $\alpha = 1$}
    
    
    
    

    \end{axis}

    \end{tikzpicture}
    }
    \caption{Beamformer directivity effect, $K=6$, $t=2$, $L=4$}
    \label{fig:sbf_plot}
\end{figure}

%% file: Figs/STYPE-Plot.tex
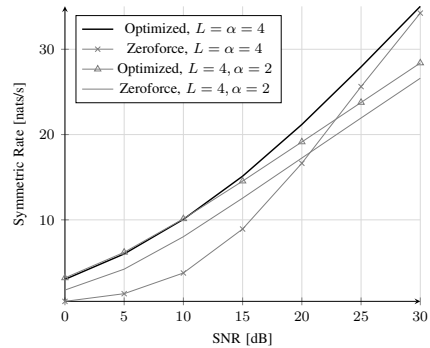
\begin{figure}[t]
    \centering
    \resizebox{0.65\columnwidth}{!}{%
    
    \begin{tikzpicture}

    \begin{axis}
    [
    axis lines = left,
    xlabel = \smaller {SNR [dB]},
    ylabel = \smaller {Symmetric Rate [nats/s]},
    ylabel near ticks,
    legend pos = north west,
    ticklabel style={font=\smaller},
    grid=both,
    major grid style={line width=.2pt,draw=gray!30},
    ]
    
    \addplot[black,thick]
    table[y=L4A4OPT,x=SNR]{Data/STYPE.tex};
    \addlegendentry{\smaller Optimized, $L=\alpha = 4$}
    
    \addplot[mark = x,gray]
    table[y=L4A4ZF,x=SNR]{Data/STYPE.tex};
    \addlegendentry{\smaller Zeroforce, $L=\alpha = 4$}
    
    \addplot[mark = triangle,gray]
    table[y=L4A2OPT,x=SNR]{Data/STYPE.tex};
    \addlegendentry{\smaller Optimized, $L=4, \alpha = 2$}
    
    \addplot[gray]
    table[y=L4A2ZF,x=SNR]{Data/STYPE.tex};
    \addlegendentry{\smaller Zeroforce, $L=4, \alpha = 2$}
    
    
    
    

    \end{axis}

    \end{tikzpicture}
    }
    \caption{Beamformer structure effect, $K=6$, $t=2$, $L=4$}
    \label{fig:stype_plot}
\end{figure}

%% file: CCGains.tex
In this section, we review the three CC-stemmed parameters in Figure~\ref{fig:gain_breakdown} that affect the finite-SNR performance of cache-aided MISO communications.


\subsection{CC-aided multiplexing effect}
CC-aided multiplexing is determined by the CC gain $t$ and indicates how much interference can be removed by the cache contents of target users. As the value of $t$ directly affects DoF, we expect the performance to improve as $t$ grows larger.

To investigate the effect of the CC-aided multiplexing gain, we use numerical simulations for a \textit{CC}-$(6,t,4,4)$ setup, where $t \in \{0,1,2\}$.
We assume the MISO-CC scheme of~\cite{shariatpanahi2018physical} is used as the baseline, and optimized beamformers are applied to maximize the performance.
The results are provided in Figure~\ref{fig:ccmux_plot} and Table~\ref{tab:tmux_bar}. Highlights of the results are:
\begin{enumerate}
    \item As expected, better performance is attained with larger $t$;
    \item Performance improvement by cC-aided multiplexing is independent of the SNR value. This is because cache-aided interference removal is also independent of SNR;
    \item Cache-aided communications is a strong tool at finite-SNR as it does not impose similar complexities of beamforming in that regime.
\end{enumerate}

\input{Figs/CCMUX-Plot}

\begin{table}[t]
    \centering
    \begin{tabular}{c||c|c|c|c|c|c|c}
         SNR ($\mathrm{dB}$) & 0 & 5 & 10 & 15 & 20 & 25 & 30  \\
         \hline \hline
         $t=2$ & 75 & 90 & 101 & 105 & 105 & 101 & 94 \\
         \hline
         $t=1$ & 48 & 53 & 57 & 61 & 62 & 60 & 56 \\
    \end{tabular}
    \caption{Rate benefit (\%) over $t=0$ case, $K=6$, $\alpha=L=4$}
    \label{tab:tmux_bar}
\end{table}

\subsection{CC-aided multicasting effect}
So far, we have only considered bit-level MISO-CC schemes where the cache-aided interference cancellation is performed \textit{after} the received signal is decoded. As mentioned earlier, another class of MISO-CC schemes with cache-aided interference cancellation \textit{before} decoding the received signal has recently gained popularity due to nice properties such as reduced subpacketization~\cite{lampiris2018adding,salehi2020lowcomplexity}, simpler beamformer design~\cite{salehi2020lowcomplexity}, and applicability to dynamic setups~\cite{salehi2021low}. However, it is also discussed in~\cite{salehi2019subpacketization,salehi2022enhancing} that these nice properties are accompanied by penalties; most notably in reduced finite-SNR performance and more complex signaling requirements in the control plane. In this paper, we investigate the former issue and show that it is related to a reduction in the coded multicasting gain. In~\cite{salehi2019subpacketization}, this gain is referred to as \textit{efficiency index}.

Let's consider a \textit{CC}-$(6,2,4,4)$ setup and review how the subpacketization can be reduced from 15 to 3 using a signal-level CC scheme.
\newline\noindent\textbf{Placement phase.} We split each file $W \in \CF$ into $P=3$ subpackets $W_1$, $W_2$, $W_3$, and store $W_1$ in users~1 and~2, $W_2$ in users~3 and~4, and $W_3$ in users~4 and~5.
\newline\noindent\textbf{Delivery phase.} Let us denote the files requested by users 1-6 as $A,B,\cdots,F$, respectively. We only need the following transmission vector:
\begin{equation}
\label{eq:signal_level_exmp}
\begin{aligned}
    \Bx_{1\cdots 6} = &A_2 \Bw_{256} + A_3 \Bw_{234} + B_2 \Bw_{156} + B_3 \Bw_{134} \\
    &+ C_1 \Bw_{456} + C_3 \Bw_{412} + D_1 \Bw_{356} + D_3 \Bw_{312} \\
    &+ E_1 \Bw_{634} + E_2 \Bw_{612} + F_1 \Bw_{534} + F_2 \Bw_{512} .
\end{aligned}
\end{equation}
Let us review the decoding process at user~1, which is interested in the first and second data terms in~\eqref{eq:signal_level_exmp}. Using~\eqref{eq:general_received_signal}, this user receives
\begin{equation*}
    y_{1\cdots 6}(1) = A_2 \Bh_1^T \Bw_{256} + A_3 \Bh_1^T \Bw_{234} + \Bh_1^T I_{BF} + \Bh_1^T I_{CC} + z_1,
\end{equation*}
where the interference terms are
\begin{equation*}
    \begin{aligned}
        I_{BF} = &B_2 \Bw_{156} + B_3 \Bw_{134} + C_3 \Bw_{412} \\
        &+ D_3 \Bw_{312} + E_2 \Bw_{612} + F_2 \Bw_{512} , \\
        I_{CC} = &C_1 \Bw_{456} + D_1 \Bw_{356} + E_1 \Bw_{634} + F_1 \Bw_{534} .
    \end{aligned}
\end{equation*}
However, the interference terms in $I_{BF}$ are nulled-out or suppressed by beamforming, and the ones in $I_{CC}$ could be regenerated and removed from $y_{1\cdots 6}(1)$ using the cached contents of user~1. So, user~1 can decode $A_2$ and $A_3$ using a SIC receiver after cache-aided interference cancellation is done in the signal domain. Comparing $\Bx_{1\cdots 6}$ in~\eqref{eq:signal_level_exmp} with its bit-level counterpart in~\eqref{eq:bit_level_exmp}, it is seen that in the signal-level scheme, we have fewer codewords and hence, need less multicasting. 

As shown in~\cite{salehi2019subpacketization}, for our example \textit{CC}-$(6,2,4,4)$ setup, different subpacketization levels of $P \in \{3,6,9,12,15\}$ are possible. Moreover, as we move from a pure signal-level scheme (smallest $P$) to a pure bit-level one (largest $P$), we can use more multicasting in the transmission. Here, we have used numerical simulations to compare the system performance for $P\in\{3,6,12,15\}$. The results are provided in Figure~\ref{fig:ccmc_plot} and Table~\ref{tab:ccmc_bar}. Highlights of the results are:
\begin{enumerate}
    \item Performance is improved as subpacketization is increased and more multicasting is supported. This is because multicasting is more power-efficient than unicasting;
    \item Performance improvement is more prominent in finite-SNR, as the rate is power-limited in this regime.
\end{enumerate}

\input{Figs/CCMC-Plot}

\begin{table}[t]
    \centering
    \begin{tabular}{c||c|c|c|c|c|c|c}
         SNR ($\mathrm{dB}$) & 0 & 5 & 10 & 15 & 20 & 25 & 30  \\
         \hline \hline
         $P=15$ & 48 & 48 & 46 & 42 & 36 & 30 & 25 \\
         \hline
         $P=12$ & 44 & 40 & 36 & 32 & 28 & 24 & 20 \\
         \hline
         $P=6$ & 10 & 10 & 10 & 11 & 11 & 10 & 9 
    \end{tabular}
    \caption{Rate benefit (\%) over $P=3$ case, $K=6$, $t=2$ ,$\alpha=L=4$}
    \label{tab:ccmc_bar}
\end{table}

\input{Figs/CCMAC-Plot}

\subsection{Superposition (MAC) effect}
In the transmission vector model in~\eqref{eq:general_transmission_vector}, each codeword $X_{\CT}$, $\CT \in \Bar{\CT}(\CS)$, includes data for every user $k \in \CT$. Let us consider a specific user $\Bar{k} \in \CS$. After the transmission of $\Bx_{\CS}$, if there exist multiple sets $\CT \in \Bar{\CT}(\CS)$ that include $\Bar{k}$, this user has to decode its requested data from a multiple access channel using a SIC receiver. However, SIC receivers are complex to implement, and hence, more effort has been recently put into designing CC schemes without a SIC requirement~\cite{mohajer2020miso,salehi2020lowcomplexity}. Interestingly, it is also shown in~\cite{mahmoodi2021low} that removing the SIC requirement could greatly simplify optimized beamformer design through iterative optimization methods.

Removing the SIC requirement (or reducing the MAC size) is possible in both bit- and signal-level schemes. However, signal-level schemes generally provide more flexibility as they are less constrained by multicasting~\cite{mohajer2020miso,salehi2020lowcomplexity}. Here, we only consider controlling the MAC size in bit-level schemes, as they enable CC-aided multicasting gain to be achieved at its full capacity. For example, let us consider a \textit{CC}-$(6,2,4,4)$ setup and its transmission vector $\Bx_{1\cdots 6}$ in~\ref{eq:bit_level_exmp}. As discussed in Section~\ref{section:misogains}, every user needs to decode its requested data from a MAC of size~10 after $\Bx_{1 \cdots 6}$ is transmitted. However, for this network, we may control the MAC size (and even avoid it altogether) by scheduling the codewords sent by $\Bx_{1 \cdots 6}$ into more transmissions. For example, to avoid the MAC, we can use the following ten transmission vectors instead:
\begin{equation}
    \begin{aligned}
        \Bx_{1 \cdots 6}^1 &= X_{123}\Bw_{456} + X_{456}\Bw_{123}, \\
        \Bx_{1 \cdots 6}^2 &= X_{124}\Bw_{356} + X_{356}\Bw_{124}, \\
        & \cdots \\
        \Bx_{1 \cdots 6}^{10} &= X_{156}\Bw_{234} + X_{234}\Bw_{156}. \\
    \end{aligned}
\end{equation}

Controlling the MAC size is first introduced in~\cite{tolli2017multi} using a $\beta$ parameter for reducing beamformer design complexity. Accordingly, we also use $\beta$ to denote the MAC size. Simulation results for the example \textit{CC}-$(6,2,4,4)$ setup are provided in Figure~\ref{fig:ccmac_plot} and Table~\ref{tab:ccmux_bar}. It is assumed that $\beta \in \{1,2,5,10 \}$, and optimized beamformers are used to maximize the performance. Highlights of the results are:
\begin{enumerate}
    \item Increasing the MAC size generally improves performance (we suspect the deviations are due to numerical errors). This is due to improved superposition coding~\cite{tolli2017multi}. However, compared with other parameters, the effect is small;
    \item Given the complexity of SIC receivers and the small gain of superposition coding, expanding the literature on MAC-avoiding bit-level CC schemes is a promising direction.
\end{enumerate}

\begin{table}[t]
    \centering
    \begin{tabular}{c||c|c|c|c|c|c|c}
         SNR ($\mathrm{dB}$) & 0 & 5 & 10 & 15 & 20 & 25 & 30  \\
         \hline \hline
         $\beta=10$ & 3 & 8 & 11 & 13 & 16 & 19 & 22 \\
         \hline
         $\beta=5$ & 11 & 12 & 12 & 12 & 14 & 17 & 18 \\
         \hline
         $\beta=2$ & 4 & 6 & 6 & 6 & 7 & 8 & 10 
    \end{tabular}
    \caption{Rate benefit (\%) over $\beta=1$ case, $K=6$, $t=2$ ,$\alpha=L=4$}
    \label{tab:ccmux_bar}
\end{table}

%% file: Figs/CCMUX-Plot.tex
\begin{figure}[t]
    \centering
    \resizebox{0.65\columnwidth}{!}{%
    
    \begin{tikzpicture}

    \begin{axis}
    [
    axis lines = left,
    xlabel = \smaller {SNR [dB]},
    ylabel = \smaller {Symmetric Rate [nats/s]},
    ylabel near ticks,
    legend pos = north west,
    ticklabel style={font=\smaller},
    grid=both,
    major grid style={line width=.2pt,draw=gray!30},
    ]
    
    \addplot[black,thick]
    table[y=t2OPT,x=SNR]{Data/CCMUX.tex};
    \addlegendentry{\smaller $t=2$}
    
    \addplot[mark = x,gray]
    table[y=t1OPT,x=SNR]{Data/CCMUX.tex};
    \addlegendentry{\smaller $t=1$}
    
    \addplot[mark = triangle,gray]
    table[y=t0OPT,x=SNR]{Data/CCMUX.tex};
    \addlegendentry{\smaller $t=0$}

    \end{axis}

    \end{tikzpicture}
    }
    \caption{Cache-aided multiplexing effect, $K=6$, $\alpha = L = 4$}
    \label{fig:ccmux_plot}
\end{figure}
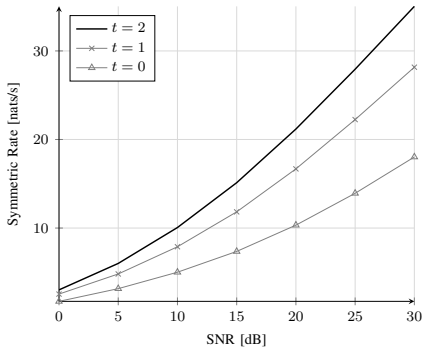

%% file: Figs/CCMC-Plot.tex
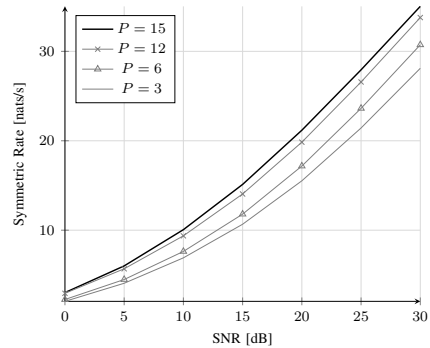
\begin{figure}[t]
    \centering
    \resizebox{0.65\columnwidth}{!}{%
    
    \begin{tikzpicture}

    \begin{axis}
    [
    axis lines = left,
    xlabel = \smaller {SNR [dB]},
    ylabel = \smaller {Symmetric Rate [nats/s]},
    ylabel near ticks,
    legend pos = north west,
    ticklabel style={font=\smaller},
    grid=both,
    major grid style={line width=.2pt,draw=gray!30},
    ]
    
    \addplot[black,thick]
    table[y=P15OPT,x=SNR]{Data/CCMC.tex};
    \addlegendentry{\smaller $P=15$}
    
    \addplot[mark = x,gray]
    table[y=P12OPT,x=SNR]{Data/CCMC.tex};
    \addlegendentry{\smaller $P=12$}
    
    \addplot[mark = triangle,gray]
    table[y=P6OPT,x=SNR]{Data/CCMC.tex};
    \addlegendentry{\smaller $P=6$}
    
    \addplot[gray]
    table[y=P3OPT,x=SNR]{Data/CCMC.tex};
    \addlegendentry{\smaller $P=3$}
    
    
    
    

    \end{axis}

    \end{tikzpicture}
    }
    \caption{Cache-aided multicasting effect, $K=6$, $t=2$, $\alpha=L=4$}
    \label{fig:ccmc_plot}
\end{figure}

%% file: Figs/CCMAC-Plot.tex
\begin{figure}[t]
    \centering
    \resizebox{0.65\columnwidth}{!}{%
    
    \begin{tikzpicture}

    \begin{axis}
    [
    axis lines = left,
    xlabel = \smaller {SNR [dB]},
    ylabel = \smaller {Symmetric Rate [nats/s]},
    ylabel near ticks,
    legend pos = north west,
    ticklabel style={font=\smaller},
    grid=both,
    major grid style={line width=.2pt,draw=gray!30},
    ]
    
    \addplot[black,thick]
    table[y=B10OPT,x=SNR]{Data/CCMAC.tex};
    \addlegendentry{\smaller $\beta = 10$}
    
    \addplot[mark = x,gray]
    table[y=B5OPT,x=SNR]{Data/CCMAC.tex};
    \addlegendentry{\smaller $\beta = 5$}
    
    \addplot[mark = triangle,gray]
    table[y=B2OPT,x=SNR]{Data/CCMAC.tex};
    \addlegendentry{\smaller $\beta = 2$}
    
    \addplot[gray]
    table[y=B1OPT,x=SNR]{Data/CCMAC.tex};
    \addlegendentry{\smaller $\beta = 1$}
    
    
    
    

    \end{axis}

    \end{tikzpicture}
    }
    \caption{Superposition (MAC) effect, $K=6$, $t=2$, $\alpha = L = 4$}
    \label{fig:ccmac_plot}
\end{figure}
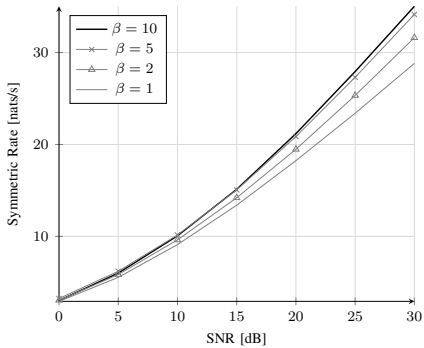

%% file: main.bbl
\begin{thebibliography}{10}
\providecommand{\url}[1]{#1}
\csname url@samestyle\endcsname
\providecommand{\newblock}{\relax}
\providecommand{\bibinfo}[2]{#2}
\providecommand{\BIBentrySTDinterwordspacing}{\spaceskip=0pt\relax}
\providecommand{\BIBentryALTinterwordstretchfactor}{4}
\providecommand{\BIBentryALTinterwordspacing}{\spaceskip=\fontdimen2\font plus
\BIBentryALTinterwordstretchfactor\fontdimen3\font minus
  \fontdimen4\font\relax}
\providecommand{\BIBforeignlanguage}[2]{{%
\expandafter\ifx\csname l@#1\endcsname\relax
\typeout{** WARNING: IEEEtran.bst: No hyphenation pattern has been}%
\typeout{** loaded for the language `#1'. Using the pattern for}%
\typeout{** the default language instead.}%
\else
\language=\csname l@#1\endcsname
\fi
#2}}
\providecommand{\BIBdecl}{\relax}
\BIBdecl

\bibitem{rajatheva2020whiteb}
\BIBentryALTinterwordspacing
N.~Rajatheva and et~al., ``{White Paper on Broadband Connectivity in 6G},''
  \emph{arXiv preprint arXiv:2004.14247}, 2020. [Online]. Available:
  \url{http://arxiv.org/abs/2004.14247}
\BIBentrySTDinterwordspacing

\bibitem{maddah2014fundamental}
M.~A. Maddah-Ali and U.~Niesen, ``{Fundamental limits of caching},'' \emph{IEEE
  Transactions on Information Theory}, vol.~60, no.~5, pp. 2856--2867, 2014.

\bibitem{mahmoodi2021non}
H.~B. Mahmoodi, M.~J. Salehi, and A.~Tolli, ``{Non-Symmetric Coded Caching for
  Location-Dependent Content Delivery},'' \emph{IEEE International Symposium on
  Information Theory - Proceedings}, vol. 2021-July, pp. 712--717, 2021.

\bibitem{salehi2022enhancing}
\BIBentryALTinterwordspacing
M.~Salehi, K.~Hooli, J.~Hulkkonen, and A.~Tolli, ``{Enhancing Next-Generation
  Extended Reality Applications with Coded Caching},'' \emph{arXiv preprint
  arXiv:2202.06814}, 2022. [Online]. Available:
  \url{http://arxiv.org/abs/2202.06814}
\BIBentrySTDinterwordspacing

\bibitem{shariatpanahi2016multi}
S.~P. Shariatpanahi, S.~A. Motahari, and B.~H. Khalaj, ``{Multi-server coded
  caching},'' \emph{IEEE Transactions on Information Theory}, vol.~62, no.~12,
  pp. 7253--7271, 2016.

\bibitem{shariatpanahi2018physical}
S.~P. Shariatpanahi, G.~Caire, and B.~Hossein~Khalaj, ``{Physical-Layer Schemes
  for Wireless Coded Caching},'' \emph{IEEE Transactions on Information
  Theory}, vol.~65, no.~5, pp. 2792--2807, 2019.

\bibitem{salehi2021MIMO}
M.~J. Salehi, H.~B. Mahmoodi, and A.~T{\"{o}}lli, ``{A Low-Subpacketization
  High-Performance MIMO Coded Caching Scheme},'' in \emph{WSA 2021 - 25th
  International ITG Workshop on Smart Antennas}, 2021, pp. 427--432.

\bibitem{tolli2017multi}
A.~T{\"{o}}lli, S.~P. Shariatpanahi, J.~Kaleva, and B.~H. Khalaj,
  ``{Multi-antenna interference management for coded caching},'' \emph{IEEE
  Transactions on Wireless Communications}, vol.~19, no.~3, pp. 2091--2106,
  2020.

\bibitem{yan2018placement}
Q.~Yan, X.~Tang, Q.~Chen, and M.~Cheng, ``{Placement Delivery Array Design
  Through Strong Edge Coloring of Bipartite Graphs},'' \emph{IEEE
  Communications Letters}, vol.~22, no.~2, pp. 236--239, 2018.

\bibitem{lampiris2018adding}
E.~Lampiris and P.~Elia, ``{Adding transmitters dramatically boosts
  coded-caching gains for finite file sizes},'' \emph{IEEE Journal on Selected
  Areas in Communications}, vol.~36, no.~6, pp. 1176--1188, 2018.

\bibitem{salehi2021low}
M.~J. Salehi, E.~Parrinello, H.~B. Mahmoodi, and A.~Tolli,
  ``{Low-Subpacketization Multi-Antenna Coded Caching for Dynamic Networks},''
  \emph{2022 Joint European Conference on Networks and Communications and 6G
  Summit, EuCNC/6G Summit 2022}, pp. 112--117, 2022.

\bibitem{mohajer2020miso}
S.~Mohajer and I.~Bergel, ``{MISO Cache-Aided Communication with Reduced
  Subpacketization},'' in \emph{IEEE International Conference on
  Communications}, vol. 2020-June.\hskip 1em plus 0.5em minus 0.4em\relax IEEE,
  2020, pp. 1--6.

\bibitem{salehi2019subpacketization}
M.~Salehi, A.~Tolli, S.~P. Shariatpanahi, and J.~Kaleva,
  ``{Subpacketization-rate trade-off in multi-antenna coded caching},'' in
  \emph{2019 IEEE Global Communications Conference, GLOBECOM 2019 -
  Proceedings}.\hskip 1em plus 0.5em minus 0.4em\relax IEEE, 2019, pp. 1--6.

\bibitem{lampiris2018resolving}
\BIBentryALTinterwordspacing
E.~Lampiris and P.~Elia, ``{Resolving a Feedback Bottleneck of Multi-Antenna
  Coded Caching},'' \emph{arXiv}, 2018. [Online]. Available:
  \url{http://arxiv.org/abs/1811.03935}
\BIBentrySTDinterwordspacing

\bibitem{salehi2020lowcomplexity}
M.~J. Salehi, E.~Parrinello, S.~P. Shariatpanahi, P.~Elia, and A.~Tolli,
  ``{Low-Complexity High-Performance Cyclic Caching for Large MISO Systems},''
  \emph{IEEE Transactions on Wireless Communications}, vol.~21, no.~5, pp.
  3263--3278, 2022.

\bibitem{mahmoodi2021low}
H.~B. Mahmoodi, B.~Gouda, M.~Salehi, and A.~Tolli, ``{Low-complexity Multicast
  Beamforming for Multi-stream Multi-group Communications},'' in \emph{2021
  IEEE Global Communications Conference, GLOBECOM 2021 - Proceedings}, 2022,
  pp. 01--06.

\end{thebibliography}
